\documentclass[12pt,epsfig]{article}
\usepackage{etex}
\usepackage{epsfig}
\usepackage{epsf,latexsym}
\usepackage{amsfonts,amssymb,amsmath} 
\usepackage{mathtools}
\usepackage{tikz}
\epsfverbosetrue
\newcommand{\double}[1]{\mathbb{#1}}

\newcommand{\rr}{\double{R}}

\newcommand{\dee}{\hbox{\rm{D}}}
\newcommand{\de}{\hbox{\rm{d}}}

\newcommand{\pa}{\partial}

\newcommand{\lb}{\left[}
\newcommand{\rb}{\right]}
\newcommand{\lp}{\left(}
\newcommand{\rp}{\right)}

\newcommand{\dpp}{\vcentcolon}
\newcommand{\bb}{\begin{eqnarray}}
\newcommand{\ee}{\end{eqnarray}}
\newcommand{\eee}{\nonumber\end{eqnarray}}
\newcommand{\qq}{\quad}

\begin{document}

\font\twelve=cmbx10 at 13pt
\font\eightrm=cmr8

\thispagestyle{empty}

\begin{center}
${}$
\vspace{3cm}

{\Large\textbf{On a weak Gau{\ss} law in general relativity and torsion}} \\

\vspace{2cm}

{\large

Thomas Sch\"ucker\footnote{thomas.schucker@gmail.com } (CPT\footnote{Centre de Physique Th\'eorique, 
Aix-Marseille Univ; CNRS, UMR 7332; Univ Sud Toulon Var;\\
13288 Marseille Cedex 9, France}),
Sami R. ZouZou\footnote{szouzou2000@yahoo.fr }
(LPT\footnote{Laboratoire de Physique Th\'eorique, D\'epartement de Physique, Facult\'e des Sciences Exactes,\\
Univ Mentouri-Constantine, Algeria})}

\vspace{3cm}

{\large\textbf{Abstract}}
\end{center}
We present an explicit example showing that the weak Gau{\ss}  law of general relativity (with cosmological constant) fails in Einstein-Cartan's theory. We take this as an indication that torsion might replace dark matter.
\vspace{3.2cm}

\noindent PACS: 04.20.-q\\
Key-Words: Einstein equation, general relativity
\vskip 2 truecm

\noindent CPT-P005-2012\\
\noindent 1203.5642\\  
\vskip 2 truecm

\section{Introduction}

Gau{\ss}' law in Maxwell's theory allows to link the total charge of an isolated charge distribution to the asymptotic strength of its electric field. In general relativity a weak form of Gau{\ss}' law holds in the static and spherically symmetric case and remains true when a cosmological constant is added. We show that this weak Gau{\ss} law fails when torsion is added \`a la Einstein-Cartan \cite{car,hehl76,tworecent}.

\section{The weak Gau{\ss} law of general relativity}

To warm up, let us quickly review the derivation of the weak Gau{\ss} law in general relativity with cosmological constant. 

{\bf In a first step} we solve the Killing equation
\bb 
\xi ^\alpha \,\frac{\pa}{\pa x^\alpha }\, g_{\mu \nu}+\,\frac{\pa \xi ^{\bar\mu }}{\pa x^\mu }\, g_{\bar\mu \nu}+\,\frac{\pa \xi ^{\bar\nu }}{\pa x^\nu }\, g_{\mu \bar\nu}=0.\label{kill}\ee
in the static, spherical case, i.e.
for the four vector fields: $\xi =\pa/\pa t$ generating time translation and 
\bb
\xi &=&-\sin\varphi \,\frac{\pa}{\pa \theta}\,  -\cos\varphi\,\frac{\cos\theta }{\sin\theta }\,  \frac{\pa}{\pa \varphi}\,  ,\\[2mm] 
\xi &=&+\cos\varphi \,\frac{\pa}{\pa \theta}\, -\sin\varphi\,\frac{\cos\theta }{\sin\theta }\,  \frac{\pa}{\pa \varphi}\, ,\\
\xi &=& \frac{\pa}{\pa \varphi}\, ,\ee
 generating the rotations around the $x$-, $y$- and  $z$-axis. In fact, since for the infinitesimal rotations, the commutator of the last two gives the first, we may discard the infinitesimal rotation around the $x$-axis.
It is a typical accountant's work to list and solve the remaining $3\times 10$ linear, first order partial differential equations for the 10 unknowns $g_{\mu \nu}$. The solution is well known: 
\bb \de\tau ^2= B\,\de t^2+2 D\,\de t\,\de r-A\,\de r^2-C\,\de \theta ^2-C\,\sin^2\theta \,\de \varphi ^2,\ee
with four functions $A,\,B,\,C$ and $D$ of $r$. By a suitable coordinate transformation we may achieve $D=0$ and $C=r^2$. Then $A$ and $B$ are positive.

{\bf In a second step} we solve the Einstein equation
\bb
\text{Ricci}_{\mu \nu}-{\textstyle\frac{1}{2}}\, \text{scalar}\,g_{\mu \nu}-\Lambda \,g_{\mu \nu}=8\pi G\,\tau_{\mu \nu},\ee
in the vacuum, $\tau_{\mu \nu}=0,$ and obtain the Kottler or Schwarzschild - de Sitter solution: 
\bb B=\,\frac{1}{A}\, = 1-\,\frac{S}{r}\, -{\textstyle\frac{1}{3}} \Lambda \,r^2,\label{kottler}\ee 
with an integration constant $S$.

{\bf In a third step} we use the weak Gau{\ss} law to determine the integration constant $S$ in terms of the total mass of a spherically symmetric mass distribution ${\tau^t}_t(r)$ with support inside a ball of radius $R$ (not to be confused with the curvature scalar, that we denote by `scalar'). This law is simply the ${^t}_t$ component of the Einstein equation:
\bb {\text{Ricci}^t}_t-{\textstyle\frac{1}{2}}\,\text{scalar} \,{g^t}_t-\Lambda \,{g^t}_t
=8\pi G\,{\tau^t}_t\,.\ee
In the static, spherical case this equation reduces to:
\bb \,\frac{1}{r^2}\,  \lb 1-\,\frac{\de}{\de r}\,\frac{r}{A}\, \rb-\Lambda =8\pi G\,{\tau^t}_t\,.\ee
Solving for $\de (r/A)/\de r$ and integrating we have:
\bb \,\frac{r}{A}\, = r-2G\int_0^r{\tau^t}_t(\tilde r)\,4\pi \tilde r^2\, \de \tilde r-{\textstyle\frac{1}{3}} \Lambda \,r^3 +K.\ee
The integration constant $K$ is seen to vanish by evaluating at $r=0$ and noting that $A(0)$ is positive. Then the interior solution is
\bb 
A(r)=\lb 1-\lp 2G\int_0^r{\tau^t}_t(\tilde r)\,4\pi \tilde r^2\, \de \tilde r\rp/r-{\textstyle\frac{1}{3}} \Lambda \,r^2\rb^{-1},\ee
and from the continuity of $A$ at $r=R$ we obtain the Schwarzschild radius $S=2GM$ with
\bb M\dpp = \int_0^R{\tau^t}_t(\tilde r)\,4\pi \tilde r^2\, \de \tilde r.\ee

\section{The weak Gau{\ss} law in Einstein-Cartan's theory}

Our task is to redo the above three steps including torsion \`a la Einstein-Cartan.

\subsection{First step: invariant connection}

 We have to solve the analogue of the Killing equation \cite{ts} for the now independent (metric) connection $\Gamma $
\bb 
\xi ^\alpha \,\frac{\pa}{\pa x^\alpha }\,  {\Gamma ^\lambda }_{\mu \nu}
-\,\frac{\pa \xi ^{\lambda  }}{\pa x^{\bar\lambda } }\,  {\Gamma ^{\bar\lambda }}_{\mu \nu}
+\,\frac{\pa \xi ^{\bar\mu }}{\pa x^\mu }\,  {\Gamma ^\lambda }_{\bar\mu \nu}
+\,\frac{\pa \xi ^{\bar\nu }}{\pa x^\nu }\,  {\Gamma ^\lambda }_{\mu \bar\nu}
+\,\frac{\pa ^2\xi ^\lambda }{\pa x^\mu \pa x^\nu}\, 
=0,\label{kill2}\ee
 for the vector fields generating time translation and rotations. These $3\times 64$ equations yield the following non-vanishing connection components:
\bb
{\Gamma ^a}_{bc}={X ^a}_{bc}(r),&& a,\,b,\,c\in \{t,r\},\\
{\Gamma ^a}_{\theta \theta }={\Gamma ^a}_{\varphi \varphi  }/\sin ^2\theta =E^a(r),&&
{\Gamma ^a}_{\theta  \varphi  }=-{\Gamma ^a}_{\varphi \theta   }=\sin \theta \,F^a(r),\\
{\Gamma ^\theta }_{a \theta   }={\Gamma ^\varphi  }_{a \varphi    }=C_a(r),&&
{\Gamma ^\theta }_{ \theta a  }={\Gamma ^\varphi  }_{ \varphi   a }=Y_a(r),\\
{\Gamma ^\theta }_{\varphi \varphi }=-\sin\theta\,\cos\theta ,&& {\Gamma ^\varphi }_{\theta \varphi }={\Gamma ^\varphi }_{\varphi \theta }={\cos\theta }/{\sin\theta} , \ee
with $8+2+2+2+2=16$ arbitrary functions $X,\, E,\,F,\,C,\,Y$ of $r$. The metricity condition,
\bb \,\frac{\pa}{\pa x^\lambda }\, g_{\mu \nu}
-{\Gamma ^{\bar\mu  }}_{\mu \lambda } g_{\bar\mu \nu}
-{\Gamma ^{\bar\nu }}_{\nu \lambda } g_{\mu\bar \nu}=0,\label{metric}\ee
reduces the 16 functions to four arbitrary functions $C_t,\,C_r,\, D_t,\,D_r$ and we have the following non-vanishing connection components:
\bb {\Gamma ^t}_{tr}={\textstyle\frac{1}{2}} {B'}/{B} ,\qq {\Gamma ^t}_{ra}=D_a\,{A}/{B}, \label{invCb}&&
{\Gamma ^r}_{rr}={\textstyle\frac{1}{2}} {A'}/{A},\qq
{\Gamma ^r}_{ta}=D_a,
\\{\Gamma ^t}_{\theta \theta  }={\Gamma ^t}_{\varphi \varphi }/\sin^2\theta= C_t\,{r^2}/{B},&&
{\Gamma ^r}_{\theta \theta  }={\Gamma ^r}_{\varphi \varphi }/\sin^2\theta=- C_r\,{r^2}/{A},\\
{\Gamma ^\theta }_{a \theta   }={\Gamma ^\varphi  }_{a \varphi    }=C_a,&&{\Gamma ^\theta }_{ \theta r  }={\Gamma ^\varphi  }_{ \varphi   r }=1/r,\\
{\Gamma ^\theta }_{\varphi \varphi }=-\sin\theta\,\cos\theta ,&& {\Gamma ^\varphi }_{\theta \varphi }={\Gamma ^\varphi }_{\varphi \theta }={\cos\theta }/{\sin\theta} .\label{invCe}
\ee

\subsection{Einstein's equation}

So far we have used a holonomic frame $\de x^\mu $. Now it will be convenient to work in an orthornormal frame $e^a=\dpp {e^a}_\mu \,\de x^\mu $. We will use the notations of reference \cite{gs}. The metric tensor reads 
$g_{\mu \nu}(x) = {e^a}_\mu(x)\,{e^b}_\nu(x)\,\eta_{ab}$. 

The connection with respect to a holonomic frame is written as a $g\ell(4)$-valued 1-form ${\Gamma^\alpha  }_\beta  =\dpp {\Gamma^\alpha  }_{\beta \mu }\,\de x^\mu $. The link between the components of the same connection with respect to the holonomic frame $\Gamma $ and with respect to the orthonormal frame $\omega $ is given by the $GL(4)$ gauge transformation with $e(x)={e^a}_\mu(x)\,\in GL(4)$;
\bb \omega =e\Gamma e^{-1}+e\de e^{-1},
\ee
or with indices:
\bb {\omega ^a}_{b \mu }={e^a}_\alpha\,  {\Gamma^\alpha  }_{\beta \mu }\,{e^{-1\,\beta }}_b +{e^a}_\alpha\,\frac{\pa}{\pa x^\mu }\,   {e^{-1\,\alpha  }}_b.\label{gaugetrans}\ee
 For  $\omega $ the metricity condition is algebraic and  means that its values ${\omega^a}_b $ are in the Lie algebra of the Lorentz group: $\omega _{ab}=-\omega _{ba}$. 
 
 In the orthonormal frame with $e={\rm diag}(\sqrt B,\, \sqrt A,\, r,\,r\sin\theta ) $, the non-vanishing components ${\omega ^a}_{b \mu }$ of the invariant connection (\ref{invCb}$-$\ref{invCe}) are:
 \bb 
 {\omega ^t}_{r t }=D_t\sqrt{A/B},&
{\omega ^t}_{r r }=D_r\sqrt{A/B},&
{\omega ^t}_{\theta \theta  }={\omega ^t}_{\varphi \varphi  }/\sin \theta =C_t\,r/\sqrt{B},\\
&
{\omega ^\theta }_{\varphi \varphi  }=-\cos\theta ,&
{\omega ^r}_{\theta \theta  }={\omega ^r}_{\varphi \varphi  }/\sin \theta =-C_r\,r/\sqrt{A}.\ee
The curvature as $so(1,3)$-valued 2-form,
\bb R\dpp=\de \omega +{\textstyle\frac{1}{2}} [\omega ,\omega ], \label{cartan1}\ee
has the following non-vanishing components,
 ${R ^a}_b=\dpp{\textstyle\frac{1}{2}} {R ^a}_{b \mu \nu }\de x^\mu \de x^\nu$:
 \bb 
{R ^t}_{r tr }=-(D_t\sqrt{A/B})',&
{R ^t}_{\theta  t\theta  }=-C_rD_t\,r/\sqrt B,&
{R ^t}_{\theta  r\theta  }=(C_t\,r/\sqrt B)'-C_rD_r\,r/\sqrt B,\nonumber\\&
{R ^t}_{\varphi   t\varphi   }=\sin\theta\, {R ^t}_{\theta  t\theta  },&
{R ^t}_{\varphi   r\varphi   }=\sin\theta \,{R ^t}_{\theta  r\theta  },\nonumber\\&
{R ^r}_{\theta  t\theta  }=C_tD_t\,r\sqrt A/B,&
{R ^r}_{\theta  r\theta  }=-(C_r\,r/\sqrt A)'+
C_tD_r\,r\sqrt A/B,\nonumber\\&
{R ^r}_{\varphi   t\varphi   }=\sin\theta \,{R ^r}_{\theta  t\theta  },&
{R ^r}_{\varphi   r\varphi   }=\sin\theta \,{R ^r}_{\theta  r\theta  },\\&&
{R ^\theta }_{\varphi   \theta \varphi   }=\sin\theta\,(1 +
C_t^2\,r^2/B-C_r^2\,r^2/A).\nonumber 
\ee
Next we compute the Ricci tensor,
\bb {{\text{Ricci}}^a}_b\dpp =\eta^{aa'}{R^c}_{a'\mu \nu}{e^{-1\mu }}_c{e^{-1\nu }}_b,\ee
whose non-vanishing components are:
\bb
{{\text{Ricci}}^t}_t&=&(D_t\,\sqrt{A/B})'/\sqrt{AB}+2\,C_rD_t/B,\\
{{\text{Ricci}}^r}_r&=&(D_t\,\sqrt{A/B})'/\sqrt{AB}+
2\,(r\,C_r/\sqrt A)'/(r\sqrt A)
-2\,C_tD_r/B,\\
{{\text{Ricci}}^t}_r&=&-2\,(C_t/\sqrt B)'/\sqrt A -2\,C_t/(r\sqrt{AB})+2\,C_rD_r/\sqrt{AB},\\
{{\text{Ricci}}^r}_t&=&-2\,C_tD_t\,\sqrt{A/B}/B,\\
{{\text{Ricci}}^\theta }_\theta &=&{{\text{Ricci}}^\varphi }_\varphi\\& =&(C_r/\sqrt A)'/\sqrt A-1/r^2+C_r/(rA)-C_t^2/B+C_r^2/A-C_tD_r/B+C_rD_t/B.\nonumber
\ee
The curvature scalar is
\bb \text{scalar}&=&{{\text{Ricci}}^t}_t+{{\text{Ricci}}^r}_r+2\,{{\text{Ricci}}^\theta }_\theta \nonumber\\
&=&
2\,(D_t\,\sqrt{A/B})'/\sqrt{AB} +4\,(C_r/\sqrt A)'/\sqrt A-2/r^2+4\,C_r/(rA)\nonumber\\
&&-2\,C_t^2/B+2\,C_r^2/A-4\,C_tD_r/B+4\,C_rD_t/B.
\ee
In an orthonormal frame the Einstein equation reads
\bb {\text{Ricci}^a}_b-{\textstyle\frac{1}{2}} \,\text{scalar}\,{\delta^a}_b-\Lambda \, {\delta^a}_b=8\pi G\,{\tau_b}^a.\ee
In our orthonormal frame $e={\rm diag}(\sqrt B,\, \sqrt A,\, r,\,r\sin\theta ) $, the energy momentum tensor $\tau_{ab} $ reads
\bb \\
{\tau^a}_b=\lp
\begin{matrix}
\rho (r)&q(r)&0&0\\
o(r)&-p_r(r)&0&0\\
0&0&-p_{\rm a}(r)&0\\
0&0&0&-p_{\rm a}(r)
\end{matrix}\rp.
\ee
The $tt,\ rr$ and $\theta \theta $ components of the Einstein equation are:
\bb
-2\,(C_r/\sqrt A)'/\sqrt A+1/r^2-2\,C_r/(rA)\qq\qq\qq\qq\qq\qq\qq\qq&&\nonumber\\
+C_t^2/B-C_r^2/A+2\,C_tD_r/B-\Lambda &=&\ \ 8\pi G\,\rho ,\label{1diagEinst}\\
1/r^2+C_t^2/B-C_r^2/A-2\,C_rD_t/B-\Lambda &=&-8\pi G\,p_r ,\\
-(D_t\,\sqrt{A/B})'/\sqrt{AB} -\,(C_r/\sqrt A)'/\sqrt A-C_r/(rA)\qq\qq\qq\nonumber&&\\
+C_tD_r/B-C_rD_t/B-\Lambda &=&-8\pi G\,p_{\rm a}.\label{3diagEinst}\ee
The two off-diagonal components read
\bb
-2\,(C_t/\sqrt B)'/\sqrt A -2\,C_t/(r\sqrt{AB})+2\,C_rD_r/\sqrt{AB}&=&8\pi G\,o\\ 
-2\,C_tD_t\,\sqrt{A/B}/B&=&8\pi 
G\,q.\ee

\subsection{Cartan's equation}

The torsion as $\rr^4$-valued 2-form,
\bb T\dpp=\dee e= \de e +\omega e,\ee
 has the following non-vanishing components, 
$T^a=\dpp {\textstyle\frac{1}{2}} T_{a'bc}\eta^{a'a}e^be^c$,
\bb
T_{ttr}=(D_t-{\textstyle\frac{1}{2}} B'/A)\,\sqrt A/B,&&
T_{rtr}=D_r/\sqrt B,\\
T_{\theta t\theta }=T_{\varphi t\varphi }=C_t/\sqrt B,&&
T_{\theta r\theta }=T_{\varphi r\varphi }=(C_r-1/r)/\sqrt A.\ee
The Cartan equation,
\bb T^ce^d\epsilon_{abcd}=-8\pi G\,s_{ab},
\ee
determines the torsion in terms of its source, the half-integer spin current. This is the Lorentz-valued 3-form $s_{ab}$ i.e. the variation of the matter Lagrangian with respect to the spin connection $\omega_{ab} $. To simplify the Cartan equation, let us 
decompose the torsion tensor  into its three irreducible parts: 
\bb T_{abc}=A_{abc}+\eta_{ab}V_c-\eta_{ac}V_b+M_{abc},\ee
with the completely antisymmetric part $A_{abc}\dpp={\textstyle\frac{1}{3}}(T_{abc}+T_{cab}+T_{bca})$, the vector part $V_c\dpp={\textstyle\frac{1}{3}} T_{abc}\eta^{ab}$, and the mixed part $M_{abc}$ characterized by $M_{abc}=-M_{acb}$, $M_{abc}\eta^{ab}=0$, and $M_{abc}+M_{cab}+M_{bca}=0$.
Likewise, we decompose the spin tensor $s_{abc}$ defined by $\ast s_{ab}=\dpp s_{abc}e^c$; 
\bb s_{abc}=a_{abc}+\eta_{ca}s_b-\eta_{cb}s_a+m_{abc},\ee
with the completely antisymmetric part $a_{abc}\dpp={\textstyle\frac{1}{3}}(s_{abc}+s_{cab}+s_{bca})$, the vector part $s_b\dpp={\textstyle\frac{1}{3}} s_{abc}\eta^{ac}$, and the mixed part $m_{abc}$ characterized by $m_{abc}=-m_{bac}$, $m_{abc}\eta^{ac}=0$, and $m_{abc}+m_{cab}+m_{bca}=0$.

Then the Cartan equation reads:
\bb A_{abc}=-8\pi G\,a_{abc},&
V_a={\textstyle\frac{1}{2}} \,8\pi G\,s_a,&
M_{cab}=-8\pi G\,m_{abc}.\ee

In the static, spherical case, we have $A_{abc}=0$, 
\bb
V_t={\textstyle\frac{1}{3}} (2\,C_t +D_r)/\sqrt B,&&
V_r={\textstyle\frac{1}{3}} (-{\textstyle\frac{1}{2}} B'/B-2/r+2\,C_r+D_t\,A/B)/\sqrt A,\ee
and
\bb
M_{rtr}={\textstyle\frac{2}{3}} (-C_t+D_r)/\sqrt B,&&
M_{trt}={\textstyle\frac{2}{3}} ({\textstyle\frac{1}{2}} B'/B-1/r+C_r-D_t\,A/B)/\sqrt A,\\
M_{\theta t\theta }=M_{\varphi t\varphi }=
-{\textstyle\frac{1}{2}}  M_{rtr} ,&&
M_{\theta r\theta }=M_{\varphi r\varphi }=
{\textstyle\frac{1}{2}}  M_{trt} .\ee

\subsection{Second step: vacuum solution}

In vacuum, $\tau_{ab}=0$ and $s_{abc}=0$, we retrieve the Kottler solution, equation (\ref{kottler}). Indeed by Cartan's equation, vanishing spin current implies vanishing torsion: $C_t=D_r=0$, $C_r=1/r$ and $D_t={\textstyle\frac{1}{2}} B'/A$ and then the invariant, metric connection (\ref{invCb} - \ref{invCe}) reduces to the (symmetric) Christoffel symbols.

\section{Third step: a Schwarzschild star with torsion}
Note that metric and connection in the static, spherical case are automatically invariant under space inversion. (This is not true in the homogeneous, isotropic case \cite{ts,st}.) However we do not have invariance under time reversal and to be precise we should say `stationary' rather than `static'.

To construct a counter example to the weak Gau{\ss} law, we do suppose invariance under time reversal. This is certainly not justified for our sun, but not unreasonable for a Schwarzschild star with constant mass density $\de \rho /\de r=0$ inside the radius $R$. The following functions are odd under time reversal and must vanish:  all connection, curvature, Ricci, energy-momentum, torsion and spin tensor components with an odd number of indices equal to $t$. Consequently $C_t,\ D_r,\ q,\ o,\ s_t$ and $m_{rtr}$ are zero, the energy momentum tensor is symmetric and we remain with four unknown functions of $r$ in the fields: $B,\ A,\ C_r$ and $D_t$.  In the sources we still have five arbitrary functions of $r$: the mass density $\rho $  the radial and azimuthal pressure $p_r,\ p_{\rm a}$  and the spin densities $s_r$ and $m_{trt}$. They define the right-hand sides of the five remaining field equations, three Einstein and two Cartan equations.

To continue, we set $D_t={\textstyle\frac{1}{2}} B'/A$  and simplify notations $C\dpp=C_r$, $s\dpp=s_r$ . Then the two Cartan equations reduce to:
\bb {\textstyle\frac{4}{3}} \,(C-1/r)/\sqrt A= 8\pi G\,s,\qq m_{trt}={\textstyle\frac{1}{2}} s.\ee
Now we may introduce a Schwarzschild star by assuming that the mass density $\rho $ and the spin density $s$ are constant, i.e. $r$-independent, with an equation of state: $s={\textstyle\frac{2}{3}} w\rho $.

Upon eliminating $C$ via the Cartan equation, the $tt$, $rr$ and $\theta \theta $ components (\ref{1diagEinst} - \ref{3diagEinst}) of Einstein's equation  reduce to:
\bb
A'/(rA^2)+1/r^2-1/(r^2A)-16\pi G\,w\rho /(r\sqrt A)
-(4\pi G\,w\rho )^2
-\Lambda &=&8\pi G\,\rho ,\label{tt}\\[2mm] 
\lb 1/(rA)+4\pi G\,w\rho /\sqrt A \rb\,B'/B
\qq\qq\qq\qq\qq\qq\qq\qq\qq\qq\qq\qq\qq\qq
&&\nonumber\\
-1/r^2+1/(r^2A)+8\pi G\,w\rho /(r\sqrt A)+(4\pi G\,w\rho )^2+\Lambda &=&8\pi G\,p_r,\\[2mm]
{\textstyle\frac{1}{2}} B''/(AB)-{\textstyle\frac{1}{4}} (A'/A+B'/B)\,B'/(AB)
\qq\qq\qq\qq\qq\qq\qq\qq\qq\qq
&&\nonumber\\
-{\textstyle\frac{1}{2}} (A'/A-B'/B)/(rA)+({\textstyle\frac{1}{2}} B'/B+1/r)\,4\pi Gw\rho /\sqrt A+\Lambda &=&8\pi G\,p_{\rm a}.
\ee
As with zero torsion, the $tt$ component decouples from the other two equations and can be integrated separately. To redo the third step of section 2 for the $tt$ component with torsion, we now need two definitions of mass: an interior mass,
\bb
M_i\dpp = \int_0^R\rho \,4\pi \tilde r^2\, \de \tilde r={\textstyle\frac{4}{3}}\pi R^3\,\rho ,\qq
\rho ={\tau^{\mu =t}}_{\nu=t}={\tau^{a =t}}_{b=t} ,\ee
and an exterior mass $M_e$ defined by the strength of the gravitational field outside, $r\ge R$,
\bb
A(r)=\dpp\lb 1-\,\frac{2G\,M_e}{r}\,-{\textstyle\frac{1}{3}} \Lambda \,r^2\rb^{-1}.\ee
In contrast to the torsionless case, they do not coincide:
\bb
M_e=M_i\lb 1+ \,\frac{6w}{R^3}\, \int_0^R\,\frac{\tilde r\,\de \tilde r}{\sqrt{A(\tilde r)}}\, +\,\frac{3w^2\,GM_i}{2R^3}\, \rb.\ee
Note that for sufficiently large $|w|$ the exterior mass exceeds the interior one, even for negative $w$. 
Note also that this mass relation depends on the interior solution $A(r)$ of the $tt$ component of the Einstein equation (\ref{tt}). This solution is not obvious (to us) and we will solve equation (\ref{tt}) numerically. As a test the numerical solution will reproduce for $w=0$ the Schwarzschild star with cosmological constant \cite{stu, boe,sch}. This solution has $p_r=p_{\rm a}=\dpp p$.  Its functions, 
\bb A=\,\frac{1}{W^2}\, ,& \qq B=(\alpha K+\beta W)^2,\qq&
p=\rho \,\left[ \,\frac{K}{\alpha K+\beta W}\, -1\right] ,\ee
are continuous at the boundary $r=R$. The auxiliary quantities used are:
\bb \gamma :={\textstyle\frac{1}{3}} (8\pi  G\,\rho +\Lambda ),&\alpha :={{\textstyle\frac{1}{2}}
 8\pi G\,\rho }/{\gamma } ,&
\beta :=({-{\textstyle\frac{1}{6}} 8\pi G\,\rho\,+\,{\textstyle\frac{1}{3}} \Lambda })/{\gamma }=1-\alpha  ,\\ 
W(r):=\sqrt{1-\gamma r^2},&
K:=W(R).&\ee

\section{Numerical solution}

Equation  (\ref{tt}) has an integrable singularity at $r=0$ which can be avoided by redefining the dependent variable, $a(r)\dpp=r/A(r)$, yielding:
\bb
1-a'-16\pi G\,w\rho \sqrt{ra}-(4\pi G\,w\rho\, r)^2-\Lambda r^2=8\pi G\rho \,r^2 .\ee
We solve this equation by a Runge-Kutta algorithm with initial condition $a(0)=0$ for $0\le r\le R$. We check that $a(r)$ remains positive for $0< r\le R$, and that $\lim_{r\rightarrow 0} {A(r)}=1$ which also ensures that 
that $r/\sqrt{A(r})=\sqrt{r\,a(r)} $ is integrable for $0\le r\le R$. Then we get the masses from
\bb M_i={\textstyle\frac{4}{3}} \pi R^3\rho ,\qq
M_e=({R-{\textstyle\frac{1}{3}}\Lambda R^3 -a(R)})/({2\,G})\, .\ee
For $w=0$ we reproduce the analytic solution $A=[1-
{\textstyle\frac{1}{3}} (8\pi  G\,\rho +\Lambda )\,r^2]^{-1}$ and have $M_i=M_e$ in accordance with the weak Gau{\ss} law. To obtain a ratio of $M_e/M_i=5$ for the sun, $M_i=M_\odot$, $R=7\cdot10^8$ m, we must choose $w=3.1$ s. We get the same ratio for a cluster, $M_i= 10^{15}\,M_\odot $, $R=3\cdot10^{23}$ m, with $w=1.33\cdot 10^{15}$ s. In these two cases the positive definite contribution ${\textstyle\frac{3}{2}} {w^2\,GM_i}/{R^3} $ to the mass ratio amounts to $6\cdot10^{-6}$ and 0.6 respectively. We have used the experimentally favoured value of $\Lambda =1.5\cdot10^{-52} \ {\rm m}^{-2}$. Setting the cosmological constant to zero however does not change the values of $M_e/M_i$ by more than $10^{-5}$.

\section{Conclusion}

A torsion induced failure of the weak Gau{\ss}  law  might be welcome with respect to some of the dark matter problems. Indeed we have already seen \cite{ts} that the Hubble diagram of super novae can be fitted by the Einstein-Cartan theory  with $w=10^{17}$ s and no  dark matter. This $w$-value is not far from the one found here for a spherical cluster. However they are far, far away from the naive microscopic  value:
\bb w=\,\frac{\hbar/2}{m_\text{proton}c^2}\, \sim 10^{-25}\ {\rm s}.\ee
It would nevertheless be interesting to compute the rotation curve of a realistic galaxy and lensing in the Einstein-Cartan theory. Both are formidable theoretical  challenges.


\begin{thebibliography}{10}
\bibitem{car} \'E.~Cartan, {\it Sur les vari\'et\'es \`a connexion affine et la th\'eorie de la r\'elativit\'e g\'en\'eralis\'ee (premi\`ere partie),} Ann.~\'Ec.~Norm.~Sup.~{\bf 40} (1923) 325.\\
{\it (premi\`ere partie, suite),} Ann.~\'Ec.~Norm.~Sup.~{\bf 41} (1924) 1.\\
 {\it (deuxi\`eme partie),} Ann.~\'Ec.~Norm.~Sup.~{\bf 42} (1925) 17.
 \bibitem{hehl76} 
 For a review see:\\
 F. W. Hehl, P. von der Heyde, G. D. Kerlick and J. M. Nester, {\it General relativity with spin and torsion: Foundations and prospects,} Rev. Mod. Phys. {\bf 48} (1976) 393.
 \bibitem{tworecent}
 Three recent reviews are:\\
 S.~Capozziello, G.~Lambiase and C.~Stornaiolo,
{\it Geometric classification of the torsion tensor in space-time,}
  Annalen Phys.\  {\bf 10 } (2001)  713.
  [gr-qc/0101038].\\
 I.~L.~Shapiro,
 {\it Physical aspects of the space-time torsion,}
  Phys.\ Rept.\  {\bf 357 } (2002)  113.
  [hep-th/0103093].\\
  M. Blagojevi\'c and F. W. Hehl (eds.), {\bf Gauge Theories of Gravitation, a
reader with commentaries} (2012) Imperial College Press, London, in press.
  \bibitem{ts}  
A. Tilquin and T. Sch\"ucker, {\it
Torsion, an alternative to dark matter?},
	arXiv:1104.0160 [astro-ph.CO],  Gen. Rel. Grav. 43 (2011) 2965.
 \bibitem{gs} M.~G\"ockeler and T.~Sch\"ucker,
{\bf Differential Geometry, Gauge Theories, and Gravity} (1987)
Cambridge Monographs on Mathematical Physics,
Cambridge University Press.
\bibitem{st}   T. Sch\"ucker and A. Tilquin, {\it
Torsion, an alternative to the cosmological constant?},
Int.\ J.\ Mod.\ Phys.\ D {\bf 21} (2012) 1250089
  [arXiv:1109.4568 [astro-ph.CO]].
	\bibitem{stu}
  Z.~Stuchl\'{i}k,
  {\it Spherically Symmetric Static Configurations of Uniform Density in
  Spacetimes with a Non-Zero Cosmological Constant,'} 
  Acta Phys.\ Slov.\  {\bf 50} (2000) 219, arXiv:0803.2530 [gr-qc].
  \bibitem{boe}
  C.~G.~Boehmer,
  {\it Eleven spherically symmetric constant density solutions with cosmological
  constant,} arXiv:gr-qc/0312027,
  Gen.\ Rel.\ Grav.\  {\bf 36} (2004) 1039.
\bibitem{sch} T. Sch\"ucker, {\it Lensing in an interior Kottler solution}, arXiv:0903.2940 [astro-ph], Gen. Rel. Grav. 42 (2010) 1991
\end{thebibliography}
\end{document}